# On the local aspect of valleytronics

Zheng-Han Huang,[1] Feng-Wu Chen,[1] and Yu-Shu G. Wu[1, 2, †]

[1] *Department of Electrical Engineering, National Tsing-Hua University, Hsin-Chu 30013, Taiwan, ROC*
[2] *Department of Physics, National Tsing-Hua University, Hsin-Chu 30013, Taiwan, ROC*

Valley magnetic moments play a crucial role in valleytronics in 2D hexagonal materials. Traditionally, based on studies of quantum states in homogeneous bulks, it is widely believed that only materials with broken structural inversion symmetry can exhibit nonvanishing valley magnetic moments. Such constraint excludes from relevant applications those with inversion symmetry, as specifically exemplified by gapless monolayer graphene despite its technological advantage in routine growth and production. This work revisits valley-derived magnetic moments in a broad context covering inhomogeneous structures as well. It generalizes the notion of valley magnetic moment for a state from an integrated total quantity to the local field called 'local valley magnetic moment' with space-varying distribution. In suitable inversion-symmetric structures with inhomogeneity, e.g., zigzag nanoribbons of gapless monolayer graphene, it is shown that the local moment of a state can be nonvanishing with sizable magnitude, while the corresponding total moment is subject to the broken symmetry constraint. Moreover, it is demonstrated that such local moment can interact with space-dependent electric and magnetic fields manifesting pronounced field effects and making possible a local valley control with external fields. Overall, a path to 'local valleytronics' is illustrated which exploits local valley magnetic moments for device applications, relaxes the broken symmetry constraint on materials, and expands flexibility in the implementation of valleytronics.

## I. INTRODUCTION

After pioneering studies on the quantum Hall effect in graphene layers [1–3], atomically thin 2D hexagonal crystals with broken inversion symmetry, e.g. gapped graphene [4–7] and monolayer transition metal dichalcogenides [8] have been recognized [9,10] to form a crucial class of topological materials with significant impacts, due to the presence of two degenerate and inequivalent band structure valleys generally designated by K and K', respectively. The valley degree of freedom has important technological implications for binary information processing and, as such, has inspired the emergence of valleytronics [11]. In addition, extensive research efforts have led to the exciting discovery of a diverse range of novel valley phenomena including valley magnetic moments [9, 10] and those connected with the moments [12–14], such as robust valley topological currents [15–21], valley-polarized interface states [15], valley-orbit [22,23] and valley Zeeman interactions [9,23], with the findings having also motivated important device proposals for valleytronic applications such as valley filters/valves [11,18,24,25], qubits [23,26–29], and FETs [30].

Traditionally, studies of valley magnetic moments have been performed from a homogeneous perspective, with important deductions specifically drawn from investigating moments of homogeneous bulk states as topological quantities [9,10]. Such a perspective has long guided the field with important influence. For instance, studies have skipped any potential nontrivial spatial dependence in the valley magnetic moment and have been focused primarily on its integrated total value. Constraints such as breaking of the structural inversion symmetry [9] have been established as rules for nonvanishing total moments and widely applied to the selection of materials in experiments and applications, with gapped AB stacked bilayer graphene [12–14] and monolayer transition metal dichalcogenides [31–33] being well-known options for experiments. Moreover, when external field-valley magnetic moment interactions are explored, primarily those between homogeneous fields and total moments have been investigated. Within the above perspective, a restricted description of the spatial dependence can in principle be provided, though. In the limit of weak, slowly-varying structural inhomogeneity, for example, such description would consist of a suitable partition of the space and application to each region the deduction drawn from the homogeneous case.

However, rigorously speaking, the quasi-homogeneous treatment of spatial dependence may overall under-describe the spectrum of valley physics, specifically that in the limit of strong inhomogeneity. From the scientific standpoint, the under-description may have overlooked interesting hidden aspects beyond the homogeneous perspective which are worthy of exploration. From the application standpoint, the under-description may raise the issue of validity concerning taking broken inversion symmetry as a universal material constraint, and clarifying such issue is critical as it impacts material options and opportunities for applications with valley magnetic moments. Inspired by both foregoing prospects, this work revisits valley-derived magnetic moments across the spectrum of inhomogeneity covering both weak and strong limits. It generalizes the notion of valley magnetic moment for a quantum state from an integrated total quantity to a local field with space-varying distribution called 'local valley magnetic moment' in the work. In suitable inversion-symmetric structures, e.g., zigzag nanoribbons of gapless graphene, where abrupt boundaries induce strong inhomogeneity, it is shown that even though the total moment of a state vanishes due to inversion symmetry, the state can nevertheless exhibit a sizable, nonvanishing local moment distribution. Moreover, it is demonstrated that such local moment can interact with space-dependent electric or magnetic fields, manifesting pronounced field effects and making possible a local valley control with external fields. Altogether, a path to 'local valleytronics' is opened up with advantages including expanded material options, among which an important one is gapless monolayer graphene. In particular, in view of available routine production with exfoliation or state-of-the-art 2D crystal growth [34–39] for such graphene, the path considerably relaxes valley-derived magnetic moment based experiments and applications.

The presentation is organized as follows. **Sec. II** discusses the notion of local valley magnetic moments in an analytical way. Specifically, it develops a current density formulation for both notional and quantitative discussions of local valley magnetic moments. In addition, it provides a compatibility discussion from the symmetry stand point, for the existence of nonvanishing local valley magnetic moments in inversion-symmetric structures. Last, interactions between local moments and magnetic and electric fields - local valley Zeeman and local valley-orbit interactions, respectively, are presented near the end of **Sec. II**. **Sec. III** performs numerical studies and presents results in connection with and validating analytical discussions in **Sec. II**. **Sec. IV** gives conclusion and outlook. **Appendix A** provides a derivation of the current density used in **Sec. II**. **Appendix B** applies the formulation developed in **Sec. II** to the calculation of local valley magnetic moments in the homogeneous bulk case. **Appendix C** provides a supplement to the compatibility discussion in **Sec. II**.

## II. LOCAL VALLEY MAGNETIC MOMENTS

For clarity, we start the discussion with graphene serving as an example, describe the notion of local valley magnetic moments in terms of an intuitive picture, and then support the picture by deriving an analytical expression of local moments in the Dirac model of graphene with inhomogeneity, which also provides in the weak inhomogeneous limit a connection with the current theoretical understanding of valley magnetic moments, as well as goes beyond the limit with an important clue given for the likely existence of nonvanishing local moments in the case of an inversion-symmetric structure. Following it, an exact, symmetry-based argument is presented to explicitly support the compatibility between foregoing existence likelihood and inversion symmetry. Last, built on these foregoing discussions, a generic, operationally defined expression of local valley magnetic moments is developed independent of materials and structures for numerical calculations.

**Figure** 1 shows a monolayer graphene crystal structure, where each unit cell consists of two carbon sites denoted by A (red) and B (blue) throughout the work. It also depicts a representative graphene electron state, in the tight-binding model including only nearest neighbor hopping and carbon atomic $2p_z$ orbitals [40] with on-site energy $\varepsilon_A = \Delta$ for the orbital on A and $\varepsilon_B = -\Delta$ for that on B. $\Delta$ is also the gap parameter characterizing the corresponding graphene band structure, with 2$\Delta$ being the gap between conduction and valence bands. In gapless graphene, $\Delta = 0$ and $\varepsilon_A = \varepsilon_B$ giving inversion symmetry between A and B sites and, thus, to the structure, too. As illustrated in **Figure 1**, the local valley magnetic moment of an electron arises out of a spin-like, local electron orbital rotation, as explained in the following. Take a near-K electron state $\phi_K$ as an example. Write the state as $\phi_{K,A} + \phi_{K,B}$ with $\phi_{K,A}$ composed of A site orbitals and $\phi_{K,B}$ B site orbitals. For a conduction (valence) band electron, the component $\phi_{K,A}$ ($\phi_{K,B}$) dominates over the other. The two components carry E″ and A″ symmetry, respectively, in the group-theoretical representation language [41], which means they carry opposite phase increments $\pm 2/3\pi$, respectively, when moving among sites. Such phase variations lead to corresponding loop currents of opposite senses (orange and blue circles, respectively) that compete with each other. Since each current is weighted by electron probability on the corresponding site, i.e., $\rho_A$ or $\rho_B$ ($\rho_{A(B)}$ = local probability on site A (B)), the competition yields a net loop current $\propto \rho_A - \rho_B$ (light green circle) and gives a 'local pseudospin' with a local magnetic moment $\propto \rho_A - \rho_B$.

To facilitate the discussion, we further introduce the term 'probability based inversion symmetry breaking' for an electron state. Irrespective of the actual situation in structural inversion symmetry, when $\rho_A = \rho_B$ ($\rho_A \neq \rho_B$), the probability based inversion symmetry is said to exist (be broken) locally in the state. Then, the local magnetic moment actually correlates with the local degree of probability based inversion symmetry breaking. For example, when $\rho_A = \rho_B$ ($\rho_A \neq \rho_B$) the moment is zero (nonvanishing) reflecting the existence (breaking) of probability based inversion symmetry.

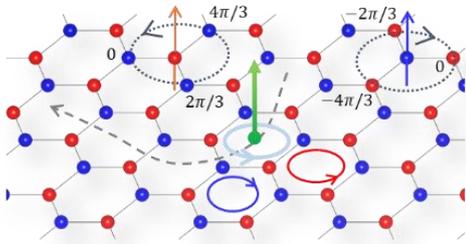

**Figure 1. Cell-orbital magnetic moment** Monolayer graphene is used for illustration. Each unit cell consists of two carbon sites (A and B). In the tight-binding model used here, on-site energy $\varepsilon_A = \Delta$ for the 2p$_z$ orbital on A site (red atom) and $\varepsilon_B = -\Delta$ for the orbital on B site (blue atom), with $\Delta = 0$ in gapless graphene. Overall, the electron performs a spin-like, local orbital rotation (light green circle) while executing a global translation (grey dashed line). Consider a near-K state, for example. Generally, it consists of the two components — $\phi_{K,A}$ composed of A site orbitals and $\phi_{K,B}$ composed of B site orbitals. The two components carry E″ and A″ symmetry in the group-theoretical representation language, with phase increments $\pm 2/3\pi$, respectively, as well as corresponding loop currents of opposite senses (orange and blue circles, respectively), resulting in the net loop current $\propto \rho_A - \rho_B$ (light green circle) and corresponding local valley magnetic moment (green, out-of-plane arrow).

Last, we make a note about how probability based inversion symmetry breaking may arise in the presence of structural inversion symmetry. Take a zigzag nanoribbon of gapless graphene for example. While the structure is invariant under structural inversion, it is well known that the bounded structure terminates on A sites on one edge and B sites on the other. Therefore, through boundary conditions on the electron state, an edge-induced AB asymmetry enters the corresponding electron probability distribution, giving distinct $\rho_A$ and $\rho_B$ and resulting in probability based inversion symmetry breaking. Further discussions about zigzag nanoribbons will be given below in this section as well as in **Sec. III**.

### 1. The Dirac model

A discussion in the Dirac model of graphene with inhomogeneity is now performed to illustrate the foregoing picture.

Consider a simple Q1D inhomogeneous structure in the absence of external fields, with the inhomogeneity derived from a spatial variation in the gap parameter of the model. For simplicity, the varying gap parameter is taken to be $\Delta(y)$ which preserves translational symmetry in the $x$ direction. Moreover, we take $\Delta(y)$ to be a regular function free of singularities and, thus, avoid complications such as those due to abrupt boundaries. Let $F$ be the Dirac two-component wave amplitude on carbon A and B sites ($F = (F_A, F_B)^t$, '$t$' = transpose), valley index $\tau = 1$ (-1) for valley K (K'), $E$ = electron energy relative to the mid-gap point, $k_x$ = wave vector relative to the Dirac point, and $(x,y)$ = cell position. $F$ satisfies the following Dirac equation ($\hbar = 1$, $e = -1$ (electron charge), and $v_F = 1$ (Fermi velocity) throughout the work) [3,23,42]:

$$H_{Dirac} F = EF,$$
$$H_{Dirac} = \begin{pmatrix} \Delta(y) & k_x - \tau \partial_y \\ k_x + \tau \partial_y & -\Delta(y) \end{pmatrix}. \quad (1)$$

Note that the usual $k_x \pm i k_y$ in the off-diagonal matrix elements of Dirac Hamiltonian for bulk graphene is now replaced by $k_x \pm \tau \partial_y$, with the substitution $k_y \to -i\partial_y$ for the structure considered here following the standard effective mass theory [43]. Eqn. (1) is a generalization of those given in References 23 and 42 for graphene ribbons to the case where $\Delta(y)$ is space varying.

From Eqn. (1), the current density operator '$j_x$' in the $x$ direction is easily constructed, giving

$$j_x = -\frac{k_x \rho}{E} - \tau \frac{\partial_y \rho_{diff}}{2E}, \quad (2)$$

where $\rho(x, y) \equiv \rho_A(x, y) + \rho_B(x, y)$, $\rho_{diff}(x, y) \equiv \rho_A(x, y) - \rho_B(x, y)$, and $\rho_{A(B)}(x, y) \equiv |F_{A(B)}(x, y)|^2$. Details of the derivation of Eqn. (2) are given in **Appendix A**. Note that both $\rho_A$ and $\rho_B$ are actually independent of $x$ due to the translational symmetry in the $x$ direction. So is $j_x$. Following the standard theory of magnetostatics [44], where the current density $\vec{j}$ in the presence of a magnetization distribution $\vec{m}$ is written as $\vec{j} = \vec{j}_{free} + \nabla \times \vec{m}$, we identify the first term in Eqn. (2) with $(\vec{j}_{free})_x$ - a free charge-composed translational current and the second term $(\nabla \times \vec{m})_x$ - a magnetization current, with the corresponding magnetization distribution $\vec{m}$ given by

$$\vec{m} = -\frac{\tau \rho_{diff}}{2E} \hat{z}. \quad (3)$$

Important implications follow Eqn. (3) as given below.

i) As $\vec{m} \propto \rho_{diff}$, it confirms the picture of local valley magnetic moments depicted in Figure 1. Moreover, a quantitative expression of local valley magnetic moment is provided by the corresponding projection '$m$', with $m \equiv \vec{m} \cdot \hat{z} = -\frac{\tau \rho_{diff}}{2E}$.

ii) For a homogeneous bulk, where $\Delta(y) = \Delta_0$, Eqn. (3) gives

$$m = \rho \, \mu_{bulk}(E, \Delta_0; \tau),$$
$$\mu_{bulk}(E, \Delta_0; \tau) \equiv -\frac{\tau \Delta_0}{2E^2}, \quad (4)$$

where $\mu_{bulk}(E, \Delta_0; \tau)$ = total valley magnetic moment of the bulk state. Derivation of Eqn. (4) is given in **Appendix B**. Eqn. (4) agrees exactly with that obtained in the traditional, homogeneous perspective with a topological, valley Berry curvature-based approach [9]. Importantly, it shows the two notable features of $\mu_{bulk}$ traditionally established within the perspective, namely, the one-to-one correspondence between $\tau$ and $\text{sgn}(\mu_{bulk})$, and vanishing $\mu_{bulk}$ in the presence of structural inversion symmetry ($\Delta_0 = 0$) [9]. Such features constitute what we call 'homogeneous perspective-based expectations or constraints.

iii) The expression of $m$ given in Eqn. (4) takes the form of a $\rho$-weighted distribution of $\mu_{bulk}$ in the $(x,y)$ space, which suggests a simple extension to the weak, slowly varying inhomogeneous case, that is, $m(x, y) = \rho(x, y) \, \mu_{bulk}(E, \Delta(x, y); \tau)$. Such a quasi-homogeneous extension would, however, subject the local moment to the rule of broken structural inversion symmetry, that is, when $\Delta(x, y) = 0$, $\mu_{bulk}(E, \Delta(x, y); \tau) = 0$ and so $m(x,y) = 0$. In contrast, the expression of $m$ given in Eqn. (3) which is suited to general inhomogeneity is less restricted. It predicts, on the contrary, the likely existence of nonvanishing $m$ when $\rho_{diff}(x, y)$ is finite, irrespective of the actual situation in structural inversion symmetry.

## 2. Local valley magnetic moments and inversion symmetry

The likely existence of nonvanishing local moments, even in the presence of structural inversion symmetry, marks an important deviation of the present local valleytronics from the traditional, homogeneous perspective based valleytronics. The likelihood can generally be argued from the standpoint of symmetry, regardless of singularities such as abrupt boundaries in structures, as follows.

Inversion-symmetric structures with translational symmetry in the *x* direction are considered. Let $m_\tau(y)$ be the local moment distribution in the transverse (*y*) dimension, for a quantum state near one of the two Dirac points with valley index $\tau$. Note that $m_\tau$ is uniform in the *x* direction, given the translational symmetry in the direction.

Firstly, we briefly apply the traditional symmetry argument [9] to the total valley magnetic moment $\int m_\tau(y)dy$ and show that it vanishes in the structure considered here. Denote $\int m_\tau(y)dy$ by M. Then an apparent conflict would come up when applying the inversion operation (Inv) as follows. 1) With the inversion being a symmetry of the structure, it follows that M remains invariant under Inv. 2) On the other hand, Inv flips the wave vector of the state and, hence, valley index, too, giving Inv($\tau$) = -$\tau$. Since valleys $\tau$ and -$\tau$ are also time reversal (TR) transforms of each other, i.e., TR($\tau$) = -$\tau$, it follows that the corresponding current loop of M reverses sense when going from $\tau$ to -$\tau$ thus leading to a sign flip in M, in conflict with the earlier conclusion of M being invariant. The conflict can only be resolved by putting M = 0.

However, the above symmetry argument does not forbid the existence of a nonvanishing $m_\tau(y)$. For example, an oscillating, antisymmetric $m_\tau(y)$, i.e., $m_\tau(-y) = -m_\tau(y)$ with nonvanishing amplitude would not violate the conclusion of vanishing M.

Below, we show the compatibility between an antisymmetric $m_\tau(y)$ and structural inversion symmetry. In **Figure 2**, such $m_\tau(y)$ is depicted in the middle graph. Applying Inv changes $m_\tau(y)$ to '$m_\tau(-y)$', with the transformed distribution shown in the left graph. On the other hand, as Inv flips the valley index as TR, it therefore changes $m_\tau(y)$ to '$m_{-\tau}(y)$' or '-$m_\tau(y)$', with the transformed distribution shown in the right graph. The agreement between the transformed Inv($m_\tau(y)$) and TR($m_\tau(y)$) demonstrates a consistency in the case of an antisymmetric $m_\tau(y)$ and, hence, concludes compatibility between such $m_\tau(y)$ and inversion symmetry. A more detailed argument is presented in **Appendix C**, in the case of zigzag nanoribbons in gapless graphene, where it provides a brief overview of state transformation under Inv and TR, and applies it to $m_\tau(y)$, as a supplement to the above discussion.

It would be worthwhile to note about the role of translational symmetry in the two examples, namely, homogeneous bulks and zigzag nanoribbons, both in gapless graphene. As already concluded, $\int m_\tau(y)dy = 0$ in both cases. But, concerning $m_\tau(y)$, a distinction resulting from the symmetry may exist between the cases, as follows. In the homogeneous bulk case, with $m_\tau$ a uniform distribution due to the translational symmetry, it is obvious that only a trivial antisymmetric distribution, i.e., $m_\tau(y) = 0$ everywhere can concur with the vanishing $\int m_\tau(y)dy$. In contrast, for inhomogeneous structures such as zigzag nanoribbons, $m_\tau(y)$ is likely space varying due to the lack of translational symmetry in the *y* direction. Therefore, even though $\int m_\tau(y)dy = 0$, it leaves plenty of room for $m_\tau(y)$ to dodge the trivial destiny if it is antisymmetric. As will be illustrated explicitly with numerical results in **Sec. III**, $m_\tau(y)$ in zigzag nanoribbons indeed oscillates with a nonvanishing amplitude as graphed in **Figure 2**.

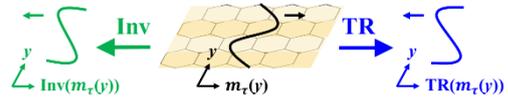

**Figure 2. Local valley magnetic moment** in a zigzag graphene nanoribbon of gapless graphene. Middle – antisymmetric distribution $m_\tau(y)$; left – transformed Inv($m_\tau(y)$); right – transformed TR($m_\tau(y)$). Thin, short horizontal arrows indicate corresponding wave vectors ($k_x$) of quantum states.

3.  **Generic definition**

A both model- and material- independent, functional derivative expression is given below to define the local valley magnetic moment in terms of the local Zeeman response to a weak probing magnetic field, as follows:

$$m(\vec{r}) = -\frac{\delta E_{Zeeman\_valley}[B_z^{(probe)}(\vec{r})]}{\delta B_z^{(probe)}(\vec{r})}\bigg|_{B_z^{(probe)}(\vec{r})=0},$$

$$E_{Zeeman\_valley}[B_z^{(probe)}(\vec{r})] = -\int m(\vec{r})B_z^{(probe)}(\vec{r})d^2r$$

(5)

($E_{Zeeman\_valley}$ = valley Zeeman energy, $B_z^{(probe)}$ = probing magnetic field). Eqn. (5) exploits the physics of local Zeeman interaction '$-m(\vec{r})B_z^{(probe)}(\vec{r})$' to operationally define $m(\vec{r})$. Without going into details, we state that it can be shown that such definition when applied to the Q1D inhomogeneous structure earlier considered in the Dirac model reproduces the same expression of *m* derived there.

Eqn. (5) can be applied to numerical studies, including those with abrupt boundaries. In the graphene case, we perform such studies with the same tight-binding model used in **Figure 1**, with the magnetic field included in the model through the Peierls substitution method [45].

In the case of a Q1D structure, $B_z^{(probe)}$ is taken to be a strip of flux as shown in **Figure 3**. Usage of the strip flux results in *m(y)* independent of *x*, consistent with translational symmetry in the *x*-direction in the structure.

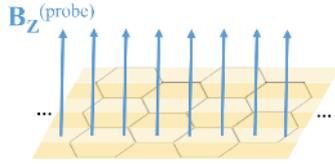

**Figure 3.** $B_z^{(probe)}$ A strip of local, vertical magnetic field is used in the case of a Q1D structure.

### 4. Effects of external fields

Interactions between local valley magnetic moments and space-dependent electric and magnetic fields are discussed below. Because derivations of the interactions are somewhat involved, the presentation below takes the following strategy. Previous results in the homogeneous bulk case are briefly mentioned, followed by conjectures for extensions to the inhomogeneous case based on the results. Rigorous results are stated at the end, with derivations given and accessible elsewhere [46].

In the homogeneous bulk case, it is known that for a bulk state the corresponding valley magnetic moment $\mu_{bulk}$ can interact with a uniform, out-of-plane magnetic field, e.g., $B_z$, shifting the state energy by the valley Zeeman term '$-\mu_{bulk}B_z$' [9,23]. In the inhomogeneous case, the foregoing result is replaced by the local expression '$-m(\vec{r})B_z(\vec{r})$', following the earlier discussion in **Sec. II** that defines the local valley magnetic moment.

Similarly, it is known that $\mu_{bulk}$ can also couple with an electric field giving rise to the valley-orbit interaction. For a bulk graphene state with wave vector $k_x$, the corresponding interaction energy is given by '$\frac{k_x}{\Delta_0}\varepsilon_y\mu_{bulk}$' [23], in the case where $\varepsilon_y$ is a uniform, in-plane electric field in the y direction. This result leads, in the Q1D case, to the conjecture of a corresponding local expression given by '$\frac{k_x}{\Delta_0}\varepsilon_y(y)m(y)$', for the interaction between the local moment *m*(y) and a space-dependent electric field $\varepsilon_y(y)$.

In the following, we restrict the attention to Q1D structures and present a rigorous statement in the linear response regime for local valley – external field interactions. Consider a quantum state with wave vector $k_x$. Let $m^{(0)}(y)$ and $E^{(0)}$ be the corresponding field-free local valley magnetic moment and electron state energy, respectively. In the linear response regime, the local valley – external field interaction energy is given by

$$E_{valley-field} = \int_{-\infty}^{\infty} \frac{k_x}{E^{(0)}} \varepsilon_y(y) m^{(0)}(y) dy$$
$$- \int_{-\infty}^{\infty} B_z(y) m^{(0)}(y) dy \quad (6)$$

with

$$\frac{k_x}{E^{(0)}} \varepsilon_y(y) m^{(0)}(y) \quad (7)$$

being the *local valley-orbit interaction* due to the electric field $\varepsilon_y(y)$ and

$$-B_z(y) m^{(0)}(y) \quad (8)$$

the *local valley Zeeman interaction* due to the magnetic field $B_z(y)$. Both interactions can serve as useful *mechanisms* for *local valley control* with space-dependent electric / magnetic fields, in analogy to their bulk counterparts which have already been demonstrated to be useful for valleytronic device applications [34]. Note that in the low energy limit where $E^{(0)} \to \Delta_0$, Eqn. (7)

reduces to the earlier conjecture '$\frac{k_x}{\Delta_0}\varepsilon_y(y)m(y)$' based on the homogeneous bulk result.

## III. NUMERICAL RESULTS

This section carries out numerical studies to illustrate i) nonvanishing local valley magnetic moments in the presence of inversion symmetry and ii) local magnetic and electric effects.

As the local magnetic moment is dominated by the difference $\rho_{diff}$, we look for structures with strong modulation on the atomic scale in order to create a pronounced contrast between A and B sites for $\rho_{diff}$ to be nonvanishing. This leads to the consideration of zigzag graphene nanoribbons, where one boundary abruptly terminates at A sites and the other at B sites thus creating a strong asymmetry between A and B sites. In all figures presented below, the same nanoribbon is studied, with the gap parameter $\Delta = 0$ eV and ribbon width $W = 65.8\ a$ ($a = 1.42\ \overset{\circ}{\text{A}}$ being the bulk lattice constant). Throughout presentations, magnetic moments are expressed in units of the Bohr magneton ($\mu_B = 5.79\times10^{-5}$ eV/Tesla).

**Figure 4** presents local valley magnetic moments of nanoribbon subband states. **(a)** shows a few valence and conduction subbands in the ribbon. States of opposite Dirac valleys are located near $k_x \sim -2.10\ a^{-1}$ and $k_x \sim 2.10\ a^{-1}$, respectively. **(b)** shows VMM$_{1/2}$ - local valley magnetic moments accumulated over half width of the ribbon (i.e., VMM$_{1/2} \equiv \int_0^{W/2} m(y)dy$), for subband states already presented in **(a)**, using the same color index scheme used in **(a)**. Note that for each subband, VMM$_{1/2}$ flips in sign for states of opposite valleys (near $k_x \sim -2.10\ a^{-1}$ and $k_x \sim 2.10\ a^{-1}$, respectively). Moreover, for each $k_x$, VMM$_{1/2}$ flips in sign, too, for corresponding conduction and valence band states related by electron-hole symmetry, for example, second conduction and valence subband states. Both flips can be attributed to the underlying time reversal symmetry and will play a role in **Figures 5** and **6** below when field effects are considered. Note that VMM$_{1/2}$ in **(b)** is sizable and can sometimes exceed 10 $\mu_B$ in the nanoribbon considered. **(c)** illustrates local valley magnetic moments (LVMMs) of second valence subband states (in the red curve shown in **(a)**) at a few selected $k_x$'s (-1.88 $a^{-1}$, -2.10 $a^{-1}$, and -2.31 $a^{-1}$) near the Dirac point at $k_x \sim -2.10\ a^{-1}$. All LVMMs shown here exhibit antisymmetry in the $y$-direction, giving vanishing total valley magnetic moments irrespective of $k_x$. **(d)** presents $\rho_A(y)$ and $\rho_B(y)$ of the second valence subband state at $k_x = -2.10\ a^{-1}$, which implies a sign oscillation in $\rho_{diff}(y)$ and, hence, in the corresponding LVMM as well in agreement with the oscillation shown by the red curve in **(c)**.

**Gapless zigzag graphene nanoribbon**

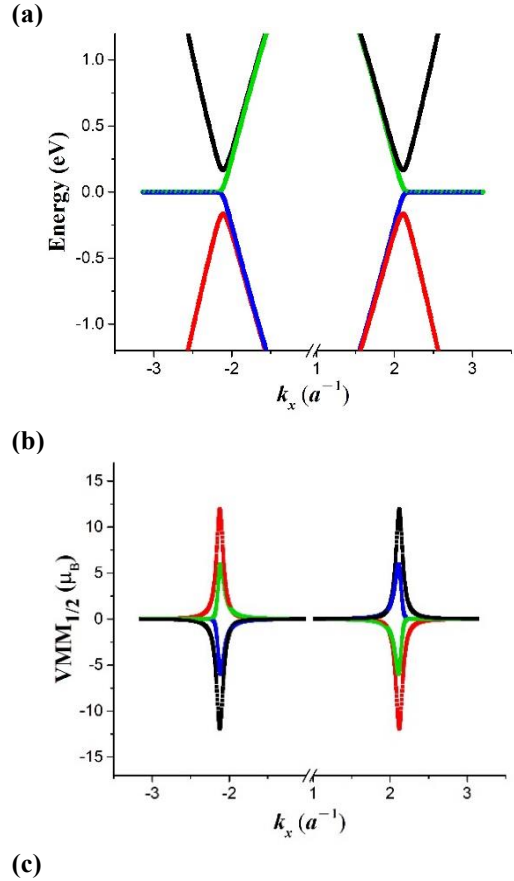

(a)

(b)

(c)

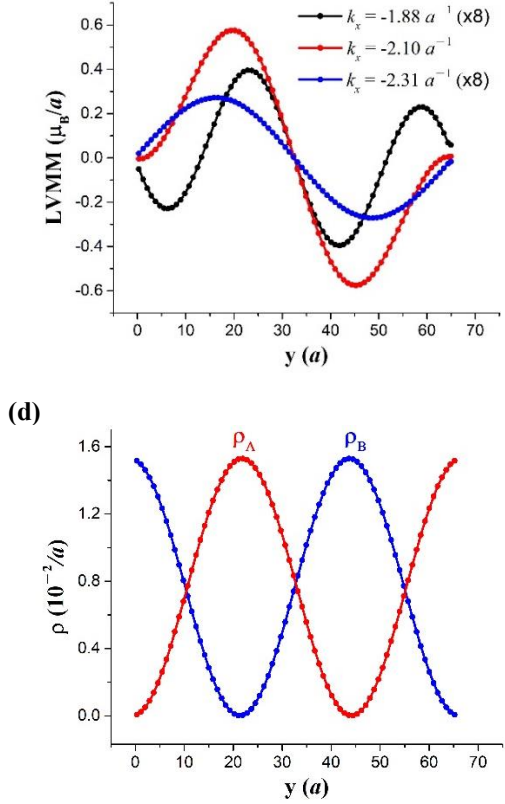

**Figure 4. Local valley magnetic moments** in a zigzag nanoribbon of gapless graphene. **(a)** shows a few nanoribbon subbands, with each subband indexed by a corresponding color. The two Dirac points are located near $k_x = -2.10\ a^{-1}$ and $k_x = 2.10\ a^{-1}$, respectively. **(b)** shows local valley magnetic moments integrated over half width of the ribbon (VMM$_{1/2}$) for subband states already presented in **(a)**, using the same color index scheme given in **(a)**. **(c)** Nontrivial, antisymmetric local valley magnetic moments (LVMMs) are obtained for second valence subband states (in the red curve shown in **(a)**) at a few selected $k_x$'s near the Dirac point with $k_x \sim -2.10\ a^{-1}$. **(d)** depicts the density distributions, $\rho_A(y)$ and $\rho_B(y)$, of the second valence subband state at $k_x = -2.10\ a^{-1}$.

**Figure 5** illustrates effects of the local valley Zeeman interaction given in Eqn. (6) and highlights the usefulness of local valley magnetic moments when total moments vanish. Introduce the magnetic field strength parameter $B_{z0}$ with $\mu_B B_{z0} = 1$ meV, which corresponds to $B_{z0} \sim 17$ Tesla. **(a)** compares the field-free subbands (black) with those when a locally varying, step-like magnetic field $B_z(y)$ is applied, with the field flux confined exclusively to the lower half ribbon, i.e., $B_z(y < W/2) = B_{z0}$ and $B_z(y > W/2) = 0$ (red). In order to interpret the graph, we apply the expression of local valley Zeeman interaction energy '$-\int_0^{W/2} B_{z0} m^{(0)}(y) dy$', which yields the product '$-B_{z0}$ VMM$_{1/2}$'. Since VMM$_{1/2}$ carries opposite signs for opposite valleys, as noted earlier in **Figure 4**, the interaction lifts the valley degeneracy resulting in the valley Zeeman splitting $\sim 17$ meV shown in the second conduction or valence subband. **(b)** compares the field-free subbands (black) with those when the whole ribbon is immersed in the uniform magnetic field given by $B_z(y) = B_{z0}$ (blue). The valley Zeeman interaction energy in this case is proportional to the total valley magnetic moment and thus vanishes. As the result, the magnetic field only induces a Landau magnetic energy shift common to both valleys without breaking the valley degeneracy.

**Local magnetic effects**

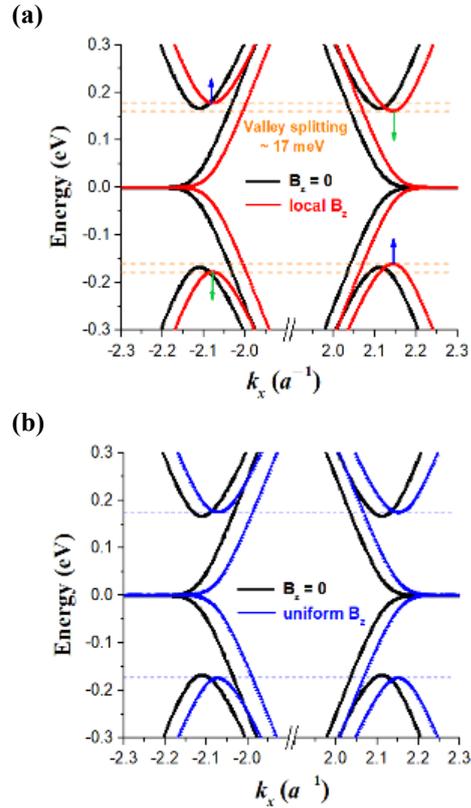

**Figure 5. Local magnetic effects** in the same zigzag nanoribbon used in **Figure 4**. **(a)** compares field-free subbands (black) with those when a locally varying, step-like magnetic field ($B_z(y)$) is applied to the lower half ribbon (red). The valley degeneracy is lifted by $B_z(y)$ leading to a valley Zeeman splitting $\sim 17$ meV. **(b)** compares field-free subbands (black) with those when a uniform magnetic field is applied (blue). It only introduces a Landau magnetic energy shift without breaking the valley degeneracy.

In **Figure 6**, we illustrate effects of the local valley-orbit interaction given in Eqn. (6). We take $\varepsilon_y(y)$ to be generated by a symmetric, piecewise linear, electrical potential $V(y)$ with the slope given by $\pm \varepsilon_{y0}$ of opposite signs for $y < W/2$ and $y > W/2$ (i.e., $\varepsilon_y(y) = -\partial_y V = \varepsilon_{y0} \, \text{sgn}(y - W/2)$). The figure compares field-free subbands (black) with those in the presence of $V(y)$ (red). In the field-free case, the subband structure has a direct gap between the second conduction and second valence bands. But in the presence of $V(y)$, band edge states of each subband shift in $k_x$ in opposite directions, for opposite valleys (near $k_x = \pm 2.10 \, a^{-1}$, respectively), giving a 'valley Rashba splitting'. Moreover, band edge states of the two subbands shift in $k_x$ in opposite directions creating a relative wave vector difference $\delta k_x \sim 0.02 \, a^{-1}$ between the two subbands' edges and, correspondingly, an indirect gap between the two subbands. Both foregoing $V(y)$-induced shifts can be explained in terms of the local valley-orbit interaction energy $\int_{-\infty}^{\infty} \frac{k_x}{E^{(0)}} \varepsilon_y(y) m^{(0)}(y) dy$, as follows. When applied to the present case, the expression reduces to '$2 \frac{k_x}{E^{(0)}} \varepsilon_{y0} \, VMM_{1/2}$', giving a linear-in-$k_x$ energy shift in the subband dispersion with a sign dependent on both the state energy $E^{(0)}$ and corresponding $VMM_{1/2}$. As noted earlier in **Figure 4**, $VMM_{1/2}$ carries opposite signs for opposite valleys as well as for second conduction and valence subbands, thus resulting in the shifts observed above. In passing, we note that due to the antisymmetry in local valley magnetic moment, a relatively simple linear potential $V(y)$ corresponding to a uniform $\varepsilon_y$ would not produce the splitting, while it would in the homogeneous bulk case with broken inversion symmetry, where it interacts with a nonvanishing $\mu_{bulk}$ and produces the splitting.

**Local electric effects**

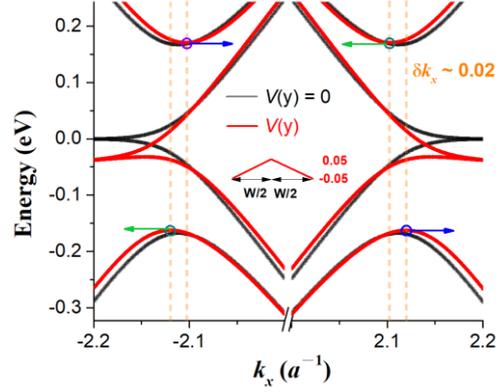

**Figure 6. Local electric effects** in the same zigzag nanoribbon used in **Figure 4**. We compare field-free subbands (black) with those in the presence of a symmetric $V(y)$ (or antisymmetric $\varepsilon_y(y)$) that varies linearly between $\pm 0.05 \, eV$ with a piecewise constant slope $\pm \varepsilon_{y0}$. In the field-free case, it shows a direct gap between the second conduction and second valence subbands. However, effects of the electric field shift various subband edges, resulting in a valley Rashba splitting as well as an indirect gap between the two subbands, with the latter characterized by a relative conduction-valence band edge wave vector difference $\delta k_x \sim 0.02 \, a^{-1}$.

In closing the section, we briefly remark on a flexibility brought by local valley – external field interactions for valley control. In the case where the local moment exhibits a sign variation in space, as illustrated in **Figure 4**, alternative local magnetic (electrical) fields with signs and distributions locally correlated to those of the local moment may produce the same valley Zeeman splitting (valley Rashba splitting) and effect the same magnetic (electric) valley control. For example, in **Figure 5 (a)**, an alternative magnetic field given by $B_z(y < W/2) = 0$ and $B_z(y > W/2) = -B_{z0}$ would produce the same valley Zeeman splitting, as can verified using the Zeeman term in Eqn. (6).

## IV. CONCLUSION

In conclusion, while being valley topological quantities, valley magnetic moments also carry a hidden dimension of local physics. As this work has shown, the presence of inhomogeneity in space makes possible distinct local valley phenomena which are not dictated by the traditional perspective developed from studies of homogeneous bulks. In order to explore the dimension, the notion of local valley magnetic moments has been introduced as a vehicle to address degrees of freedom beyond total valley magnetic moments. An operational definition is given to such moments in terms of local magnetic response.

Both analytical and numerical analysis have been performed for local valley magnetic moments giving interesting findings as summarized below. In graphene, for example, the local valley magnetic moment is shown to be tied to the local site probability difference '$\rho_A - \rho_B$', which suggests the breaking of local probability-based inversion symmetry, in place of structural inversion symmetry, as the condition for existence of and applications based on valley-derived magnetic moments. By relaxing the structural inversion symmetry constraint on materials, the study has expanded the family of valleytronic materials. In particular, it adds to the list gapless monolayer graphene, an important material which is relatively accessible, for magnetic moment-based experiments and applications. In addition, the local valley magnetic moment variable introduced is also application-suited, as it is directly linked to local valley- external field interactions. Specifically, where total valley magnetic moments vanish, local valley Zeeman and local valley-orbit interactions have been shown to exist and manifest pronounced magnetic and electric effects, respectively. Such effects can be exploited for local valley control and provide a conduit to 'local valleytronics' for the implementation of valleytronics.

Last but not the least, the novel local valley phenomena revealed suggest the exciting direction of *valley engineering* - design and search for inhomogeneous structures to tailor local valley physics for applications.

## ACKNOWLEDGMENT

We thank Yen-Ju Lin for technical support in numerical calculations. We acknowledge the financial support of MoST, ROC through Contract No. MOST 110-2112-M-007-038. F.-W. C. acknowledges partial support from Academia Sinica.

† Corresponding author. Email: yswu@ee.nthu.edu.tw

## APPENDIX A. CURRENT DENSITY

Consider a simple Q1D inhomogeneous structure exhibiting translational symmetry in the *x* direction. The Dirac Eqn. (1) is expanded giving rise to the following:

$$\Delta(y)F_A + (k_x - \tau\partial_y)F_B = EF_A$$
$$(k_x + \tau\partial_y)F_A - \Delta(y)F_B = EF_B \quad . \tag{A1}$$

Multiplying the two equations in Eqn. (A1) by $F_B^*$ and $F_A^*$, respectively, and taking complex conjugates of the resultant equations, we obtain the following:

$$\Delta(y)F_B^*F_A + F_B^*(k_x - \tau\partial_y)F_B = EF_B^*F_A$$
$$-\Delta(y)F_A^*F_B + F_A^*(k_x + \tau\partial_y)F_A = EF_A^*F_B$$
$$\Delta(y)F_B F_A^* + F_B(k_x - \tau\partial_y)F_B^* = EF_B F_A^*$$
$$-\Delta(y)F_A F_B^* + F_A(k_x + \tau\partial_y)F_A^* = EF_A F_B^* \tag{A2}$$

Combining the four wave equations in Eqn. (A2), we obtain

$$2k_x\rho + \tau\partial_y \rho_{diff} = 2Ej_x^{particle}, \tag{A3}$$

where $j_x^{particle} = F_A^*F_B + F_B^*F_A$ is the particle current density in the Dirac model [3]. This gives the charge current density

$$j_x = -\frac{k_x\rho}{E} - \tau\frac{\partial_y \rho_{diff}}{2E} \tag{A4}$$

shown in Eqn. (2).

## APPENDIX B. VALLEY MAGNETIC MOMENT IN THE BULK

In the homogeneous bulk case, the Dirac equation is given by [3]

$$H_{Dirac}F = EF,$$
$$H_{Dirac} = \begin{pmatrix} \Delta_0 & k_x - i\tau k_y \\ k_x + i\tau k_y & -\Delta_0 \end{pmatrix}, \tag{B1}$$

with the following solution

$$F_A = \rho^{1/2} (k_x - i\tau k_y)/[k^2 + (E-\Delta_0)^2]^{1/2},$$
$$F_B = -\rho^{1/2} (\Delta_0 - E)/[k^2 + (E-\Delta_0)^2]^{1/2}, \tag{B2}$$

where *E* is the electron energy given by $(k^2 + \Delta_0^2)^{1/2}$.

The substitution of Eqn. (B2) into the expression of *m* in Eqn. (3) yields

$$m = -\frac{\tau\rho_{diff}}{2E} = -\frac{\tau\rho[k^2 - (\Delta_0 - E)^2]}{2E[k^2 + (\Delta_0 - E)^2]}, \tag{B3}$$

which can be transformed by straightforward mathematics into the form of Eqn. (4) using the energy dispersion $E = (k^2 + \Delta_0^2)^{1/2}$ to express $k^2$ in terms of *E* and $\Delta_0$.

# APPENDIX C. STATE TRANSFORMATION UNDER Inv AND TR

We give an overview of state transformation under Inv and TR for zigzag nanoribbons in gapless graphene with translational symmetry in the $x$ direction, and then apply it to the local moment.

For valley $\tau$, a nanoribbon state satisfies the following Dirac equation

$$H_{Dirac} F = EF,$$
$$H_{Dirac} = \begin{pmatrix} 0 & k_x - \tau\partial_y \\ k_x + \tau\partial_y & 0 \end{pmatrix}, \quad (C1)$$
$$F_A(W/2) = F_B(-W/2) = 0.$$

The last line in Eqn. (C1) provides the boundary condition on $F$ [42]. For the discussion below, we shall denote the corresponding local valley moment of $F$ by $m_\tau(y)$.

Under Inv, $k_x \to -k_x$ and $\tau \to -\tau$, so Eqn. (C1) transforms to the one below

$$H_{Dirac}^{(Inv)} F^{(Inv)} = EF^{(Inv)},$$
$$H_{Dirac}^{(Inv)} = \begin{pmatrix} 0 & -k_x + \tau\partial_y \\ -k_x - \tau\partial_y & 0 \end{pmatrix}, \quad (C2)$$
$$F_A^{(Inv)}(W/2) = F_B^{(Inv)}(W/2) = 0,$$

with $F^{(Inv)}$ given by

$$F_A^{(Inv)}(y) = -F_B(-y),$$
$$F_B^{(Inv)}(y) = F_A(-y), \quad (C3)$$

$F^{(Inv)}$ given above satisfies both the transformed Dirac equation and the boundary condition as can be easily verified. As indicated by $F^{(Inv)}$ in Eqn. (C3), Inv switches A and B sites and at the same time induces the mirror reflection $y \to -y$. The site switch effectively flips the valley index of the state and offsets the previous valley flip in the Dirac Hamiltonian. Overall, with only the reflection in effect, it results in

$$\text{Inv}(m_\tau(y)) = m_\tau(-y). \quad (C4)$$

Under TR, again $k_x \to -k_x$ and $\tau \to -\tau$, so Eqn. (C1) becomes also one for valley '$-\tau$' given by

$$H_{Dirac}^{(TR)} F^{(TR)} = EF^{(TR)},$$
$$H_{Dirac}^{(TR)} = \begin{pmatrix} 0 & -k_x + \tau\partial_y \\ -k_x - \tau\partial_y & 0 \end{pmatrix}, \quad (C5)$$
$$F_A^{(TR)}(W/2) = F_B^{(TR)}(-W/2) = 0,$$

with the solution given by

$$F_A^{(TR)}(y) = F_A^*(y) = F_A(y),$$
$$F_B^{(TR)}(y) = -F_B^*(y) = -F_B(y). \quad (C6)$$

Above, we have used the fact that for the zigzag nanoribbon bound state, $F_A$ and $F_B$ can be taken to be real. As TR produces only a valley flip here, we obtain

$$\text{TR}(m_\tau(y)) = m_{-\tau}(y) = -m_\tau(y). \quad (C7)$$

Last, as Eqns. (C2) and (C5) are identical, with the assumption that the solutions for a given $k_x$ are nondegenerate, indeed as numerical shown in **Figure 4 (a)**, we conclude that $F^{(Inv)}$ and $F^{(TR)}$ describe the same state leading to $\text{Inv}(m_\tau(y)) = \text{TR}(m_\tau(y))$, that is, $m_\tau(-y) = -m_\tau(y)$. We thus conclude that $m_\tau(y)$ is antisymmetric in zigzag nanoribbons of gapless graphene.